\documentclass[5p,twocolumn,times]{elsarticle}

\usepackage{lineno,hyperref}
\modulolinenumbers[5]

\journal{Physics Letters B}

\usepackage{slashbox}

\begin{document}

\begin{frontmatter}

\title{
Non-coherent character of isoscalar pairing 
probed with Gamow-Teller strength: 
New insight into $^{14}$C dating $\beta$ decay
}

\author[jaea,cns]{Yutaka Utsuno\corref{cor1}}
\ead{utsuno.yutaka@jaea.go.jp}
\author[rcnp,osaka]{Yoshitaka Fujita}

\cortext[cor1]{Corresponding author}
\address[jaea]{Advanced Science Research Center, Japan Atomic Energy
Agency, Tokai, Ibaraki 319-1195, Japan}
\address[cns]{Center for Nuclear Study, University of Tokyo, Hongo,
Bunkyo-ku, Tokyo 113-0033, Japan}
\address[rcnp]{Research Center for Nuclear Physics, Osaka University, 
Ibaraki, Osaka 567-0047, Japan}
\address[osaka]{Department of Physics, Osaka University, Toyonaka, 
Osaka 560-0043, Japan}

\begin{abstract}
We investigate the phase coherence of isoscalar pairs from  
the $B({\rm GT};0^+_1\,T\!=\!1 \to 1^+_1\,T\!=\!0)$ values 
in two-particle 
configurations 
of $A=6$, 18, and 42 nuclei 
and two-hole configurations 
of $A=14$ and 38 ones. 
We find that these Gamow-Teller (GT) matrix elements 
are always constructive and thus enlarged 
under isovector- and isoscalar-pairing Hamiltonians, whereas 
the observed GT strengths are strongly hindered for the two-hole 
configurations, including the famous $^{14}$C dating $\beta$ decay. 
This indicates 
that the actual isoscalar pair, unlike the isovector pair, 
has no definite phase coherence, which can work against 
forming isoscalar-pair condensates.
\end{abstract}


\end{frontmatter}



\section{Introduction}
Pairing correlation is one of the 
most basic properties 
widely seen in quantum many-body problems including condensed-matter 
physics and nuclear physics. 
This is quite a common phenomenon caused by attractive interactions 
between constituent particles. In nuclei, the source of the attraction 
is short-range nucleon-nucleon forces, owing to 
which time-reversal pairs with large spatial overlap gain much energy 
and then a condensate of the Cooper pairs \cite{cooper56pr} occurs. 
Whereas isovector (IV) pairing (like-particle pairing) with $(J,T)=(0,1)$
is firmly established for instance 
by extra binding energies in even-even nuclei, 
a condensate of 
isoscalar (IS) proton-neutron pairs with $(J,T)=(1,0)$ 
appears quite elusive \cite{frauendorf14ppnp}. 
This is puzzling because mean attraction in the IS channel 
is much stronger than in the IV channel.  
Possible signals for IS-pairing correlation
have been explored 
for instance in terms of binding energies, rotational responses,  
Gamow-Teller (GT) $\beta$-decay properties, and proton-neutron transfer 
amplitudes 
(see \cite{frauendorf14ppnp} for a review). 
Of particular interest is that IS-pairing correlation is 
predicted to be quite 
sensitive to double-$\beta$ decay matrix elements 
\cite{vogel86prl, engel88prc, hinohara14prc, menendez16prc,engel16arxiv}. 

A condensate of pairs is based on the formation of an energetically 
stable pair in two-particle configurations. 
Using a simple attractive force, 
Cooper has shown that the binding 
energy of the lowest-energy pair is much larger than the ones of the other
eigenstates and also than the scale of 
two-body matrix elements \cite{cooper56pr}. 
This happens because a coherent combination of paired 
configurations---with a specific combination of 
signs---cooperatively 
works to lower energy \cite{heyde}. 
Whether such a coherent IS pair is formed 
in nuclei may provide a key to elucidating the origin of elusive 
IS pairing, 
but much attention has not been paid to phase coherence.

In this Letter, 
we show that the IS-pairing interaction always causes a specific 
combination of signs in the lowest $(J,T)=(1,0)$ state 
for any two-particle ($2p$) configuration and for any  
two-hole ($2h$) one and that the resulting 
$B({\rm GT};0^+_1\,T\!=\!1 \to 1^+_1\,T\!=\!0)$ value is enhanced. 
While this property well accounts for 
the low-energy super GT state for $A=6$, 18, and 42 \cite{fujita14prl}, 
it fails to explain strongly hindered $B({\rm GT})$ values for the $2h$ 
configurations, including the famous $^{14}$C dating $\beta$ decay.
This is a clear signature that the IS pair in reality 
does not take any definite signs,  
in contrast to what occurs for the ideal IS pairing. 
The IS pairing is thus fragile in nature,  
constituting an essential difference from the IV pairing which 
always favors definite signs and is thus robust. 


\section{GT strength in $2p$ and $2h$ configurations}

We start with 
an overview of observed GT strengths in $2p$
and $2h$ configurations on top of the $LS$-closed shells. 
Hereafter we restrict ourselves to the initial and final 
states with the quantum numbers $(J_i, T_i, T_{iz})=(0, 1, \pm 1)$ 
and $(J_f, T_f, T_{zf})=(1, 0, 0)$, respectively.
In Table~\ref{tab:bgt}, experimental 
$B({\rm GT};0^+_1 \to 1^+_{1,2})$ 
values are summarized for the $p$, $sd$, and $pf$ shells. 
For the $2p$ configurations, one can clearly see that the 
$B({\rm GT})$ is concentrated in the $1^+_1$ state 
for any valence shell considered, 
which is named 
the low-energy super GT state \cite{fujita14prl}. 
In contrast, the $2h$ configurations have a striking difference: 
most of the GT strength  
is exhausted by excited states, especially the $1^+_2$ state. 
It is noted that 
the $A=14$ case is well known for radiocarbon dating, which utilizes 
the very long half-life of $^{14}$C, $5730\pm 30$ yr, to determine 
the age of organic materials. 

\renewcommand{\thefootnote}{\fnsymbol{footnote}}
\begin{table}[t]
\caption{Experimental $B({\rm GT};0^+_1 \to 1^+_{1,2})$ values
 for $2p$ and $2h$ configurations in the $p$, $sd$, and $pf$ shells. Data 
taken from \cite{fujita15prc,ensdf}.}
\label{tab:bgt}
\begin{tabular}{cccccc}
 \hline
 & \multicolumn{3}{c}{$2p$} & \multicolumn{2}{c}{$2h$} \\
 \cline{2-4} \cline{5-6}
 & $p$ & $sd$ & $pf$ & $p$ & $sd$\\ 
 & $A=6$ & $A=18$ & $A=42$ & $A=14$ & $A=38$\\ \hline
 1$^+_1$ & 4.7 & 3.1 & 2.2 & 
 $3.5\times10^{-6}$\,\footnotemark
 & 0.060 \\
 1$^+_2$ & & 0.13 & 0.10 & 2.8 & 1.5\\ \hline
\end{tabular}
\end{table}
\footnotetext{For the $^{14}$C$\to ^{14}$N $\beta$
 decay. The corresponding $B({\rm GT})$ value for 
 the $^{14}$O$\to ^{14}$N $\beta$ decay is $2.0\times10^{-4}$.}

We examine how this strong asymmetry in GT strength 
between the $2p$ and $2h$ configurations arises 
in the framework of the shell model. 
The case of the $p$ shell is now taken as an example, but 
similar discussions are applicable to other shells. 
In the shell model, 
$2p$ and $2h$ configurations can be treated in a unified 
way in terms of particle-hole conjugation, since 
particle-particle two-body matrix elements are 
identical with the corresponding hole-hole matrix elements
\cite{talmi}. 
The only difference between $2p$ and $2h$
configurations concerning the GT transition 
is single-particle energies. 
Keeping this in mind, we 
calculate 
the GT matrix elements by changing 
$\Delta\varepsilon_p = \varepsilon(p_{1/2})-\varepsilon(p_{3/2})$, 
where $\varepsilon$ stands for the single-particle energy.
The values of $\Delta\varepsilon_p$ 
for the $2p$ ($A=6$) and $2h$ ($A=14$) configurations 
are $0.1$~MeV and $-6.3$~MeV, respectively, taken from 
the CKII interaction \cite{cohen67npa}.

\begin{figure}[t]
 \begin{center}
 \includegraphics[width=7.5cm,clip]{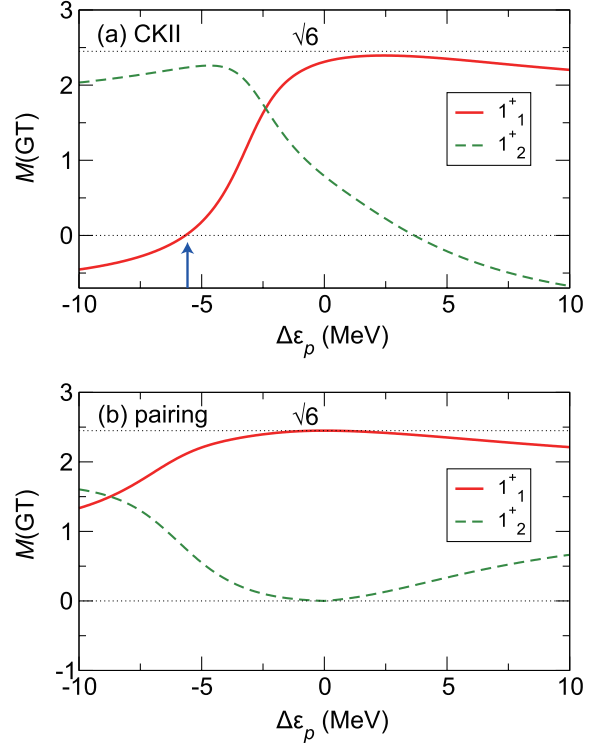}
 \caption{
(color online). Calculated $M({\rm GT};1^+_{1,2})$
for two-nucleon configurations in the $p$ shell 
as a function of $\Delta\varepsilon_p$. 
No quenching factor is used. 
(a) The CKII interaction \cite{cohen67npa} 
and (b) the IV- and IS-pairing interactions 
are used. 
}
 \label{fig:gt}
 \end{center}
\end{figure}

For the two-body part, we first use the CKII interaction as 
a realistic one, and show 
the calculated GT matrix elements 
$M({\rm GT};1^+_{k})=\langle 1^+_{k}|| \sigma t^{\pm} ||
0^+_1 \rangle$ ($k=1,2$) in Fig.~\ref{fig:gt}~(a).
When the initial and final 
states are expanded as 
$|0^+_k \rangle =\sum_{ab} \alpha^{\rm IV}_{ab}(k)
|ab J_i T_i\rangle$ 
and 
$|1^+_k \rangle =\sum_{ab} 
\alpha^{\rm IS}_{ab}(k)|abJ_f T_f\rangle$, respectively, 
$M({\rm GT};1^+_{k})$ is  
expressed by the sum of single-particle contributions as 
\begin{equation}
\label{eq:gt_decomp}
M({\rm GT};1^+_k) = 
\sum_{abcd} m_{abcd}(1^+_k),
\end{equation}
where $m_{abcd}(1^+_k) = {\alpha^{\rm IS}_{ab}}^{\ast}(k)
\alpha^{\rm IV}_{cd}(1)
\langle ab J_f T_f || \sigma t^{\pm} ||
cd J_i T_i \rangle$. 
For $\Delta\varepsilon_p > 0$, the $M({\rm GT};1^+_1)$ value 
strongly enhances, well reproducing the observed GT strength 
for $A=6$. 
This quantity is very close to the sum-rule limit of  
$\sqrt{6}\simeq 2.45$
up to $\Delta\varepsilon_p \simeq 5$~MeV, and then 
gradually decreases to the $p_{3/2}$ single-particle 
limit of $\sqrt{10/3}\simeq 1.83$. 
For $\Delta\varepsilon_p < 0$, on the other hand, 
the $M({\rm GT};1^+_1)$ value 
sharply decreases with decreasing $\Delta\varepsilon_p$. 
It crosses the $M({\rm GT})=0$ line  
at $\Delta\varepsilon_p =-5.7$~MeV, thus accounting for 
the vanishing GT strength observed in the $\beta$ 
decay of $^{14}$C. 

\section{GT strength with pairing interactions}
To probe 
pairing properties in the $2p$ and $2h$ configurations, 
it is interesting to compare those realistic shell-model calculations 
to the ones using the IS- and IV-pairing interactions. 
The IV- and IS-pairing interactions are equivalent 
to the $L=0$ part of the surface delta interaction (SDI)
in the $LS$ coupling, hence the simplest interaction of short-range 
central-force character. 
The IV- and IS-pairing interactions are defined as 
$V^{\rm IVpair}=G^{\rm IV}\sum_{\mu} P^{\dag}_{\mu}P_{\mu}$ and 
$V^{\rm ISpair}=G^{\rm IS}\sum_{\mu} D^{\dag}_{\mu}D_{\mu}$, where 
$P_{\mu}^{\dag} = \sqrt{1/2}\sum_{nl} (-1)^{l} \sqrt{2l+1} 
[a^{\dag}_{nl} a^{\dag}_{nl}]^{L=0,S=0,T=1}_{M_L=0, M_S=0, M_T=\mu}$ and 
$D_{\mu}^{\dag} = \sqrt{1/2}\sum_{nl} (-1)^{l} \sqrt{2l+1} 
[a^{\dag}_{nl} a^{\dag}_{nl}]^{L=0,S=1,T=0}_{M_L=0, M_S=\mu, M_T=0}$.
Those $jj$-coupled
two-body matrix elements 
are expressed with the Condon-Shortley phase convention
as \cite{evans81npa,poves98plb}
\begin{eqnarray}
 \langle a b J T \left| V^{\rm IVpair} \right| c d J T \rangle 
& = & G^{\rm IV}\chi^{\rm IV}_{ab}\,\chi^{\rm IV}_{cd}\,
\delta_{J0}\,\delta_{T1}, \label{eq:iv_mat}\\
\langle a b J T \left| V^{\rm ISpair} \right| c d J T \rangle
& = & G^{\rm IS}\chi^{\rm IS}_{ab}\,\chi^{\rm IS}_{cd}\,
\delta_{J1}\,\delta_{T0}, \label{eq:is_mat}
\end{eqnarray}
with
\begin{eqnarray}
\chi^{\rm IV}_{ab} &=& (-1)^{l_a}\sqrt{j_a+1/2}\,\delta_{ab} 
\label{eq:chi_iv}\\ 
\chi^{\rm IS}_{ab} &=& \sqrt{\frac{2}{1+\delta_{ab}}}
(-1)^{j_a-1/2}\sqrt{(2j_a+1)(2j_b+1)} \nonumber \\
& & \times \left\{ \begin{array}{ccc}
	      1/2 & j_a & l_a \\
              j_b & 1/2 & 1
	      \end{array}\right\} \delta_{n_a n_b}\delta_{l_a l_b} 
\label{eq:chi_is}, 
\end{eqnarray}
where $a$, $b$, $c$, and $d$ stand for single-particle states  
with quantum numbers $(n_a, l_a, j_a)$ {\it etc}., and 
$\delta_{ab}$ is the abbreviation for 
$\delta_{n_a n_b}\delta_{l_a l_b}\delta_{j_a j_b}$. 
The strengths 
$G^{\rm IV}$ and $G^{\rm IS}$ are negative for attractive 
interactions, and here we set $G^{{\rm IV}}=-3.4$~MeV 
and $G^{\rm IS}=-2.8$~MeV so that 
$\sum_{ab\ne cd} |\langle ab | V^{\rm IVpair} | cd \rangle_{J=0,T=1}|^2$
and $\sum_{ab\ne cd} |\langle ab |V^{\rm ISpair} | cd \rangle_{J=1,T=0}|^2$
become those of the CKII interaction. 

The results of the above pairing interactions  
are plotted in Fig.~\ref{fig:gt}~(b). 
Similar to the CKII interaction, 
the enhancement of the GT matrix element 
from the single-particle limit occurs 
for $\Delta\varepsilon_p > 0$. 
However, the trend for $\Delta\varepsilon_p < 0$ is 
completely different. 
The $M({\rm GT};1^+_1)$ value decreases  
rather mildly as $\Delta\varepsilon_p$ moves away from zero. 
This is a monotonic decrease which asymptotically approaches 
the $p_{1/2}$ single-particle limit of $\sqrt{2/3}\simeq$0.82 
and never vanishes. 

The essential feature of the pairing interaction shown in
Fig.~\ref{fig:gt}~(b) is that the $M({\rm GT};1^+_1)$ value enlarges  
compared to the $p_{3/2}$ and $p_{1/2}$ single-particle limits for 
$\Delta\varepsilon_p > 0$ and $\Delta\varepsilon_p  <0$, respectively. 
This is caused by the constructive interference of 
$m_{abcd}(1)$ in Eq.~(\ref{eq:gt_decomp}), 
and such an in-phase character has been presented 
for the $pf$-shell case of $^{42}$Ca 
\cite{fujita14prl, fujita15prc, bai14prc} on the basis of numerical 
analyses using the shell model and the random-phase approximation. 
It is still not very clear, however, why the constructive interference 
occurs and whether it is realized for different valence shells. 

In order to answer this question, we find a theorem 
concerning the sign of $m_{abcd}(1^+_1)$. 
\newtheorem{thm}{Theorem}
\begin{thm}
\label{thm1}
 {\it All} of the 
$m_{abcd}(1^+_1)$ values are of the same sign for 
{\it any} valence shell and for {\it any} 
single-particle splitting when the two-body 
matrix elements are given by the pairing interactions 
of Eqs.~(\ref{eq:iv_mat}) and (\ref{eq:is_mat})
with negative $G^{\rm IV}$ and $G^{\rm IS}$. 
\end{thm}

\newproof{pot}{Proof}
\begin{pot}
We first consider the signs of the two-body matrix elements of 
Eqs.~(\ref{eq:iv_mat}) and (\ref{eq:is_mat}). 
Since the sign of $\chi^{\rm IV}_{ab}$ is $(-1)^{l_a}$ 
[see Eq.~(\ref{eq:chi_iv})], 
all the matrix elements of $V^{\rm IVpair}$ can be negative (or zero) when 
one takes 
a phase convention $| \overline{a b J_i T_i} \rangle = 
(-1)^{l_a}| a b J_i T_i \rangle$. 
Similarly, for the IS pairing one can easily show that 
the sign of  $\chi^{\rm IS}_{ab}$ is $(-1)^{j_b-1/2}$,
thus obtaining entirely negative matrix elements of $V^{\rm ISpair}$
with a phase convention 
$| \overline{a b J_f T_f} \rangle = 
(-1)^{j_b-1/2}| a b J_f T_f \rangle$. 
Hereafter we refer to those phase choices as pairing 
phase convention, and the components of the eigenvectors 
in this convention are expressed by $\bar{\alpha}^{\rm IV}_{ab}(k)$ 
and $\bar{\alpha}^{\rm IS}_{ab}(k)$. 

Thus, when the IS- and IV-pairing interactions are taken, 
their off-diagonal Hamiltonian matrix elements 
in the pairing phase convention 
are completely negative or zero for any two-nucleon configuration,  
since single-particle energies do not change 
the off-diagonal matrix elements 
in the $jj$-coupling. 
For such matrices having non-positive off-diagonal matrix elements,  
it is generally true that 
all the components of the 
lowest eigenvector are of the same sign  
according to a version of the Perron-Frobenius theorem
in linear algebra\footnote{One can easily prove this case 
by showing the expectation value of 
$\vec{v}=(+\alpha_1, \ldots, +\alpha_k, -\alpha_{k+1}, \ldots, -\alpha_n)$ 
is greater than or equal to that of 
$\vec{v}'=(+\alpha_1, \ldots, +\alpha_k, +\alpha_{k+1}, \ldots, +\alpha_n)$ 
for any $\alpha_i\ge 0$.}. 
We thus obtain 
$\bar{\alpha}^{\rm IV}_{ab}(1)\ge 0$ and 
$\bar{\alpha}^{\rm IS}_{ab}(1)\ge 0$ for any $(a, b)$. 
As shown in Table~\ref{tab:mgt}, the GT matrix elements 
between two-nucleon wave functions satisfy 
$\langle \overline{ab J_f T_f} || \sigma t^{\pm} ||
\overline{cd J_i T_i} \rangle \le 0$
for any $(a,b)$ and $(c,d)$ concerned, 
hence the same sign of $m_{abcd}(1^+_1)$.
\end{pot}

\begin{table}[t]
\caption{GT matrix elements 
$\langle f|| \sigma t^{\pm} || i \rangle$ in the pairing phase
convention, 
where the two-nucleon wave functions $i$ ($T=1$) and $f$ ($T=0$) are 
denoted as $(ab)$ by using $j_>=l+1/2$ and $j_<=l-1/2$. 
We present only the basis states $i$ and $f$ that appear with 
the pairing Hamiltonians. 
}
\label{tab:mgt}
\begin{tabular}{c|cccc}
\hline
\backslashbox{$i$}{$f$}
& $(j_> j_>)$ & $(j_> j_<)$ & $(j_< j_>)$ & $(j_< j_<)$ 
\\ \hline
$(j_> j_>)$ & $-\sqrt{\frac{2(2l+3)}{2l+1}}$ & 
$-2\sqrt{\frac{l}{2l+1}}$ & $-2\sqrt{\frac{l}{2l+1}}$ & 0 \\
$(j_< j_<)$ & 0 & $-2\sqrt{\frac{l+1}{2l+1}}$ &
$-2\sqrt{\frac{l+1}{2l+1}}$ & $-\sqrt{\frac{2(2l-1)}{2l+1}}$ \\
\hline
\end{tabular}
\end{table}

This is a mathematically exact statement, 
and therefore 
provides a robust basis for the occurrence of 
the low-energy super GT state \cite{fujita14prl} in $2p$
configurations. 

\section{Phase coherence in the IS pair}
As indicated 
by the above proof, 
the key to obtaining  
the constructive interference of $m_{abcd}(1^+_1)$ is 
that all the off-diagonal Hamiltonian matrix elements are of 
the same sign which causes phase coherence 
in the IS and IV pairs. 
We stress that only the signs are relevant. 
The simplest case of the phase coherence is found in 
the original paper of the Cooper pair \cite{cooper56pr}, 
where all the off-diagonal matrix elements 
between paired electrons 
near the Fermi surface are taken to be $-|F|$ and 
all the diagonal matrix elements are zero.  
In this case, the lowest eigenvector is 
$(\alpha_1, \alpha_2, \ldots, \alpha_n)=(1, 1, \ldots, 1)/\sqrt{n}$, 
and the corresponding energy eigenvalue is $-(n-1)|F|$. 
The enhancement of eigenenergy compared to the off-diagonal 
matrix elements, called pairing gap, is due to the phase 
coherence. 
In nuclei it is well known that such a coherent pair is 
formed between like-particles. 
All the off-diagonal $(J,T)=(0,1)$ matrix elements are indeed negative 
in realistic shell-model Hamiltonians of 
CKII, USD \cite{brown88usd}, KB3 \cite{poves81pr}
and GXPF1 \cite{honma02prc}. 
Similar phase coherence in the IS pair is expected to be formed 
on the basis of the IS-pairing Hamiltonian, giving rise to 
the constructive interference of $m_{abcd}(1^+_1)$. 
In reality, however, 
the $2h$ configurations have nearly vanishing $B({\rm GT})$ values 
as shown in Table~\ref{tab:bgt}, 
pointing to destructive interference. 
Hence, the coherent IS pairs are not always formed 
with realistic interactions 
because of the opposite sign in some of the $(J,T)=(1,0)$ two-body matrix 
elements. 

\begin{figure}[b]
 \begin{center}
 \includegraphics[width=8.0cm,clip]{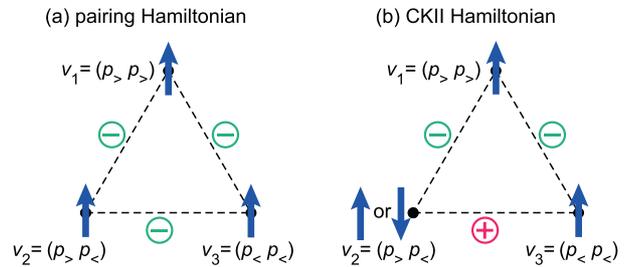}
 \caption{(color online). Graphical illustration of the signs of $(J,T)=(1,0)$ 
off-diagonal two-body matrix elements (represented as $\oplus$ 
and $\ominus$) in the pairing phase convention 
and the resulting signs of the lowest eigenstate.  
The $p$-shell cases using (a) the IS pairing Hamiltonian and 
(b) the CKII Hamiltonian are compared. 
The up and down arrows stand for positive and negative 
coefficients of $|v_i\rangle$, respectively.
}
 \label{fig:fr}
 \end{center}
\end{figure}

Taking the $p$ shell as an example, 
we present in Fig.~\ref{fig:fr} an intuitive picture about 
the formation of coherent and non-coherent IS pairs 
for the (a) IS-pairing and (b) CKII Hamiltonians, respectively, where 
the pairing phase convention is used. 
Now the lowest eigenstate is expressed as $\sum_i \alpha_i |v_i\rangle$ 
by using the basis states $|v_i\rangle$.
Before proceeding to detailed discussions, 
it should be reminded that the negative sign of an off-diagonal 
matrix element between two basis vectors $|v_i\rangle$ 
and $|v_j\rangle$, denoted as $h_{ij}$, 
favors the same sign of $\alpha_i$ and $\alpha_j$ 
and a positive $h_{ij}$ favors the opposite sign 
in the lowest eigenstate. 
Here we mean $|v_1\rangle=|\overline{p_{>}p_{>}}\rangle$, 
$|v_2\rangle=|\overline{p_{>}p_{<}}\rangle$ and 
$|v_3\rangle=|\overline{p_{<}p_{<}}\rangle$, where 
$p_>$ and $p_<$ are $p_{3/2}$ and $p_{1/2}$, respectively.
As illustrated in Fig.~\ref{fig:fr} (a), 
the obtained coherent pair is quite stabilized by the IS-pairing 
Hamiltonian, since any combination of 
$(i,j)$
satisfies the above rule. 
On the other hand, the CKII Hamiltonian has a positive $h_{23}$
and negative $h_{12}$ and $h_{13}$. 
In this case, there must be at least a combination of 
$(i,j)$
that does not comply with the above rule, as illustrated in 
Fig.~\ref{fig:fr}~(b). 
This is analogous to the geometrical frustration in magnetism 
\cite{frustration}, 
although the physical situation is rather different. 
The signs of $\alpha_i$ can no longer be uniquely determined, 
and the actual signs depend on the diagonal terms. 
For the $A=6$ system, the values of  
$h_{11}$, $h_{22}$ and $h_{33}$ are close to one another, 
and then the same sign of $\alpha_1$ and 
$\alpha_2$ is realized because of 
$|h_{12}|\gg |h_{23}|\simeq |h_{13}|$. In this case 
$|v_3\rangle$ has a small amplitude of the opposite sign, hence 
contributing little to the eigenstate. 
The dominance of $|v_1\rangle$ and $|v_2\rangle$ of the same 
sign accounts for the enhanced $B({\rm GT})$ value. 
For the $A=14$ system, $h_{11}$ is higher than $h_{33}$ 
by more than 10~MeV, so that the ground state is dominated by 
$|v_2\rangle$ and $|v_3\rangle$. The resulting signs of 
$\alpha_2$ and $\alpha_3$ are opposite 
because of $h_{23}>0$, thus leading to the nearly vanishing $B({\rm GT})$ 
value. 

In this way 
the favorable signs for $|a b{\:J\!=\!0\:T\!=\!1}\rangle$ 
in the IS pair 
are not definite and depend on the core assumed 
for realistic interactions whose 
off-diagonal $(J,T)=(1,0)$ matrix elements are not 
completely of the same sign 
in the pairing convention. 
This is an essential difference between IV and IS 
pairing, and clearly works {\it against} forming an IS-pair 
condensate. 

We point out that non-coherent IS pairs can be 
probed with pair-transfer strength, which is regarded as 
a good measure of IS pairing correlation
\cite{frauendorf14ppnp}. 
The IS-pair creation and removal strengths are now defined as 
$|\langle J_f T_f ||| D^{\dag} ||| J_i T_i \rangle|^2$ 
and $|\langle J_f T_f ||| D ||| J_i T_i \rangle|^2$, respectively, 
and we consider the transition to the lowest $(J_f,T_f)=(1,0)$ state. 
By using the CKII interaction, the IS-pair removal strength
from $^{16}$O is only $5.3\times 10^{-3}$, 
while the IS-pair creation strength
on $^{4}$He is 8.1. 
A similar strong asymmetry between the IS-pair creation
and removal is also obtained for the $sd$ shell. 

\begin{figure}[t]
 \begin{center}
 \includegraphics[width=7.5cm,clip]{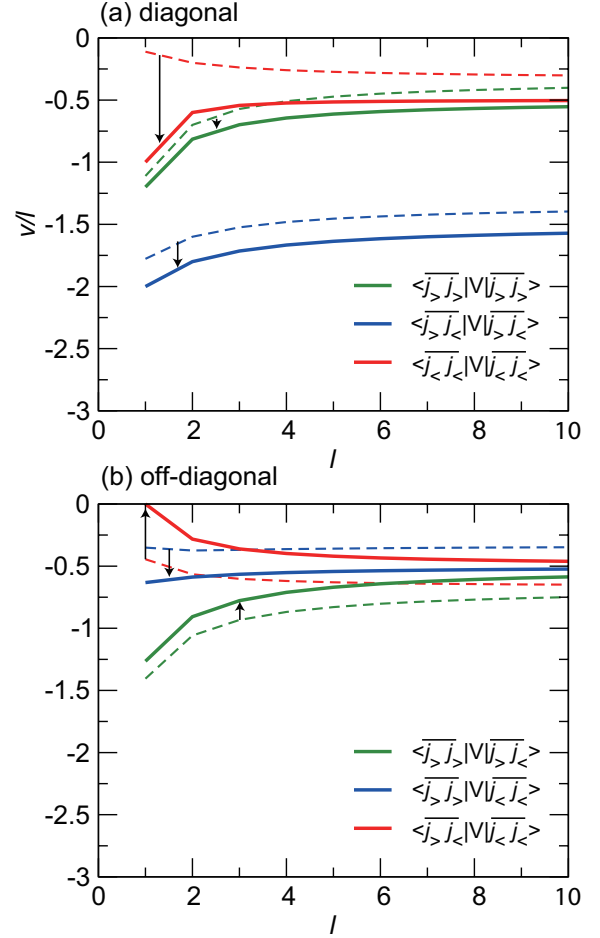}
 \caption{(color online). 
(a) Diagonal and (b) off-diagonal matrix elements of 
$\langle \overline{ab} | V | \overline{cd} \rangle_{J=1, T=0}$ 
devided by $l$ for the SDI, where 
the dashed and solid lines stand for the $L=0$ and total matrix elements, 
respectively. 
The strength of the interaction is determined so that its $L=0$ term can 
be the same as the IS-pairing interaction with $G^{\rm IS}=-1$. 
}
 \label{fig:sdi}
 \end{center}
\end{figure}

Finally, we briefly survey the origin of difference in the signs 
of the 
off-diagonal $(J,T)=(1,0)$ matrix elements 
between realistic interactions and 
the IS-pairing interaction. 
First we consider the central forces. 
Since the IS-pairing Hamiltonian is equivalent to 
the $(L,S)=(0,1)$ term of the SDI, 
the dominance of the $L=0$ central force is the source of the coherent 
IS pairs, as well as for the usual IV pairing. 
While the $(L,S)=(2,1)$ term is absent in the $(J,T)=(0,1)$
matrix elements, this term can modify the $(J,T)=(1,0)$ matrix elements. 
In Fig.~\ref{fig:sdi}, 
the effect of the $(L,S)=(2,1)$ term 
is presented for various orbital angular momenta $l$ 
by using the SDI. 
In general, 
$\langle \overline{j_> j_>}|V| \overline{j'_> j'_<}\rangle$ 
and $\langle \overline{j_< j_<}|V| \overline{j'_> j'_<}\rangle$ 
matrix elements due to $L=2$ are positive with the SDI, 
and strong cancellation between $L=0$ and 2 occurs 
especially for 
$\langle \overline{j_< j_<}|V| \overline{j'_> j'_<}\rangle$ 
with low $l$ and $l'$. For $l=1$, for instance, 
$\langle \overline{j_> j_>}|V| \overline{j_> j_<}\rangle$ 
still has a large negative value, whereas 
$\langle \overline{j_< j_<}|V| \overline{j_> j_<}\rangle$ 
vanishes. 
In addition, 
the $L=2$ term gives rise to the opposite sign between 
$\langle \overline{j j}|V| \overline{j'_> j''_<}\rangle$ 
and 
$\langle \overline{j_> j_<}|V| \overline{j'_> j''_<}\rangle$ 
for $l'=l''-2$ and $n'=n''+1$, 
thus causing frustration among $|\overline{j j}\rangle$, 
$|\overline{j_> j_<}\rangle$ and $|\overline{j'_> j''_<}\rangle$ 
when $\langle \overline{j j}|V| \overline{j_> j_<} \rangle <0$ 
is satisfied. 
We confirm that finite-range interactions lead to essentially 
similar results to the SDI 
by using the $V_{\rm MU}$ interaction \cite{otsuka10prl}, 
but some quantitative differences appear. 
For instance, the 
$\langle \overline{p_< p_<}|V| \overline{p_> p_<}\rangle$ 
matrix element, which is exacly zero for the SDI, 
is positive by taking only the $(S,T)=(1,0)$ term 
of the  $V_{\rm MU}$, 
and this matrix element can be positive or negative depending on the strength 
of the $(S,T)=(0,0)$ term. 
It should be noted that the $(S,T)=(0,0)$ term vanishes for 
zero-range interactions. 

Another important source to change the signs of $(J,T)=(1,0)$ 
matrix elements is the non-central forces. It has been pointed out 
by Jancovici and Talmi that 
phenomenological tensor forces are needed to account for the extraordinary 
long lifetime of $^{14}$C \cite{jancovici54pr}. It is worth mentioning 
that its microscopic origin has recently been discussed 
from {\it ab initio} approaches 
\cite{aroua03npa, holt08prl, holt09prc, maris11prl, ekstrom14prl}. 

In the present context, the $^{14}$C lifetime problem is a manifestation 
of the non-coherent IS pair formed by 
the positive sign of 
$\langle \overline{p_> p_<} | V | \overline{p_< p_<} \rangle$ 
due to the $L\ne 0$ central forces and 
the non-central forces. This idea can readily be 
applied to other cases. 
For instance, 
the small $B({\rm GT})$ value for $A=38$ (see Table~\ref{tab:bgt}) 
is caused by the positive sign of 
$\langle \overline{d_> d_<} | V | \overline{d_< d_<}\rangle$ 
made in a similar way to 
$\langle \overline{p_> p_<} | V | \overline{p_< p_<} \rangle>0$.
In contrast, the $\langle \overline{j_> j_>} | V | \overline{j_> j_<} 
\rangle$ matrix elements have always large negative values both 
in schematic and realistic interactions, 
thus causing the low-energy 
super GT state
\cite{fujita14prl} in $2p$ configurations.


\section{Conclusion} 
We have shown that the strong asymmetry in the 
$B({\rm GT};0^+_1 \to 1^+_1)$ values between $2p$ and $2h$ 
systems
is a clear signature that a coherent combination of 
$(J,T)=(1,0)$ pairs is not necessarily formed. 
By introducing the pairing phase convention and the idea of
frustration, we have presented a comprehensive but mathematically robust 
explanation as to why the ideal IS-pairing interaction always leads to phase 
coherence regardless of single-particle energies 
but realistic interactions do not. 
This is in sharp contrast to the IV pairing in realistic interactions, 
and may provide a key to elucidating the origin of elusive 
IS-pair condensates in nature. 
It is of great interest to investigate how modern microscopic effective 
interactions predict the $(J,T)=(1,0)$ matrix elements in a wide 
range of nuclear shells and how 
the non-coherent effect changes observables in more complex nuclei, 
including double-$\beta$ decay matrix elements. 

\section*{Acknowledgement}
Y.U. thanks M.~Mori for fruitful discussions and N.~Shimizu 
for his careful reading of the manuscript. 
This work was supported in part by JSPS KAKENHI, under Grant
No. JP15K05094 and No. JP15K05104.


\end{document}